\documentstyle[aps,12pt]{revtex}
\begin{document}

\newcommand{\pslash}{p\kern-0.4em/}
\newcommand{\Aslash}{A\kern-0.4em/}

\renewcommand{\theequation}{\arabic{section}.\arabic{equation}}
\renewcommand\b[1]{\mbox{\boldmath$#1$}}

\title{Fermion Determinants in Static, Inhomogeneous Magnetic Fields}
\author{M.P. Fry \thanks{Electronic address: mpfry@maths.tcd.ie}}
\address{School of Mathematics,\\
         Trinity College,\\
         Dublin 2,\\
         Ireland.}
\maketitle
\begin{abstract}
The renormalized fermionic determinant of QED in 3 + 1 dimensions,
$\mbox{det}_{{ren}}$, in a static, unidirectional, inhomogeneous
magnetic field
with finite flux can be calculated from the massive Euclidean
Schwinger model's determinant, $\mbox{det}_{{Sch}}$, in the same field by
integrating $\mbox{det}_{{Sch}}$, over the fermion's mass.
Since $\mbox{det}_{{ren}}$ for
general fields is central to QED, it is desirable to have
nonperturbative information on this determinant, even for the
restricted magnetic fields considered here.  To this end we continue
our study of the physically relevant determinant $\mbox{det}_{{Sch}}$.  It is
shown that the contribution of the massless Schwinger model to
$\mbox{det}_{{Sch}}$ is cancelled by a contribution from the massive sector of
QED in 1 + 1 dimensions and that zero modes are suppressed in
$\mbox{det}_{{Sch}}$.  We then
calculate $\mbox{det}_{{Sch}}$ analytically in the presence of a finite flux,
cylindrical magnetic field.  Its behaviour for large flux and small
fermion mass suggests that the zero-energy bound states of the
two-dimensional Pauli Hamiltonian are the controlling factor in the
growth of $\ln \mbox{det}_{{Sch}}$.  Evidence is presented that
$\mbox{det}_{{Sch}}$ does
not converge to the determinant of the massless Schwinger model in the
small mass limit for finite, nonzero flux magnetic fields.\\

PACS number(s): 11.15.Tk, 12.20. Ds

\end{abstract}

\section{Introduction}
\setcounter{equation}{0}

Fermion determinants produce an effective measure for the boson fields
of a Euclidean field theory when the fermions are integrated out, as
shown long ago by Matthews and Salam [1].  Such determinants in an
external boson field, or a random boson field with a cutoff, are
infinite dimensional and need to be defined [2] before the boson
fields can be integrated.  Once defined, their analysis is notoriously
difficult, especially if they possess a symmetry that should be
preserved.  For example, most classical estimates of fermion
determinants [3] invariant under a local U(1) transformation violate
this invariance.  Lack of nonperturbative information on fermion
determinants is reflected in the necessity to make loop expansions or
the more extreme quenched (valence) approximation.  As a result physical
effects predicted by the theory may be lost.\\

Nonperturbative information on the fermion determinant, such as its
growth in the complex coupling plane, is central to an analysis of the
nature of the perturbation series of the associated field theory [4].
Intuition tells one that Fermi statistics, visible in the
alternating signs of the determinant's loop expansion, ought to slow
down the growth of a perturbation series with order.\\

There is also the question of stability.  Specifically, in the case of
an Abelian gauge field, the measure is Gaussian, so that if the fermion
determinant grows faster than an inverted Gaussian, it is doubtful that it is
integrable respect to the gauge field's measure.\\

Given the sparseness of nonperturbative information on physically
relevant fermion determinants in four dimensions we thought it useful
to try to find a solvable example for a broad class of boson fields.
An obvious choice is the fermion determinant of quantum
electrodynamics in time-independent, unidirectional, inhomogeneous
magnetic fields.  Although of physical interest, this choice of fields
suffers from the fact that they are a set of measure zero as far as
the functional integral over the vector potential is concerned.
Nevertheless, the fermion determinant for such magnetic fields
remains unsolved except for the special case of a homogeneous magnetic
field that was dealt with over half a century ago by Euler,
Heisenberg, and Weisskopf [5] and later on again by Schwinger [6].
A thorough understanding of this problem would be helpful for a more
general understanding of the physical content of fermion determinants
in quantum electrodynamics.  As we will see below, there are some
significant simplifications that recommend this problem for
analysis.\\

This paper is organised as follows.  In Sec. II we review previous
relevant results.  Section III justifies our restricted choice of
magnetic fields on which the results of Sec. II rely.  We then go on in
Sec. IV to discuss the suppression of zero modes in the fermion
determinant of the massive Schwinger model.  As Sec. II makes clear,
the four-dimensional determinant is obtained by integrating this
determinant over the fermion's mass.  Section IV also illustrates the
profound change in the two-dimensional determinant when the fermions
are given a mass.  In Sec. V we calculate the massive Schwinger model's
determinant for a finite, nonzero flux magnetic field.  Section VI
contains a discussion of the zero-mass limit of the massive Schwinger
model's determinant, and Sec. VII summarizes our results.\\

\section{Review of Previous Results}
\setcounter{equation}{0}

In a previous paper [7] it was shown that the renormalized fermion
determinant, $\mbox{det}_{{ren}}$, of quantum electrodynamics in four
dimensions
(QED$_4$) in a smooth, static, unidirectional magnetic field with fast
decrease at infinity is related to the Euclidean, massive Schwinger
model's (Euclidean QED$_2$'s) determinant, $\mbox{det}_{{Sch}}$, in the
same magnetic field.  Specifically,

\begin{eqnarray}
\ln \mbox{det}_{{ren}} &=& \displaystyle\frac{e^2V_{||}}{4 \pi^2}
\displaystyle\int \displaystyle\frac{d^2k}{(2\pi)^2} |\hat{B} (k)|^2
\displaystyle\int^1_0 dz z (1 - z)\ln \left[ \displaystyle\frac{k^2z(1
- z) + m^2}{m^2}\right] \nonumber \\
&& \quad + \displaystyle\frac{V_{||}}{2\pi}
\displaystyle\int^\infty_{m^2} d M^2 \ln \mbox{det}_3(M^2),
\end{eqnarray}
where the gauge invariant $\mbox{det}_3$ is related to $\mbox{det}_{{Sch}}$ by

\begin{equation}
\ln \mbox{det}_{{Sch}} = -\displaystyle\frac{e^2}{2 \pi}
\displaystyle\int \displaystyle\frac{d^2 k}{(2 \pi)^2} |\hat{B}(k)|^2
\displaystyle\int^1_0 dz \displaystyle\frac{z(1 - z)}{k^2 z(1-z) +
m^2} + \ln \mbox{det}_3 (m^2).
\end{equation}
Since the first term in (2.2) is the contribution to the effective
action from the second-order vacuum polarization graph, $\ln \mbox{det}_3$
may be viewed as the sum of all one-loop fermion graphs in two
dimensions beginning in fourth order.  Our designation of this sum by
$\ln \mbox{det}_3$ follows the definition of Seiler [2] and Simon [3], with
the third-order graph vanishing by $C$-invariance.  Equation (2.1) states
that the sum of the corresponding graphs in QED$_4$ is obtained by
integrating $\ln \mbox{det}_3$ over the fermion mass $m$.  The first term in
(2.1) is the contribution to the effective action of QED$_4$ from the
on-shell renormalized second-order vacuum polarization graph.  The
function $\hat{B}$ is the Fourier transform of the magnetic field,
which may be asumed to point along the $z$-axis, in which case
$V_{||}$ is the volume of the $zt$ box.  In both (2.1) and (2.2) the
determinants are defined by Schwinger's proper time definition [6].
Note that the charge, $e$, will always occur in the combination $eB$ in
position space,which has the invariant dimension $L^{-2}$.  Note also
that in Euclidean QED$_2$ potentials associated with unidirectional
magnetic fields are a set of measure one.\\

The lesson of (2.1) is that the massive Schwinger model [8] is more
than a model in view of its direct bearing on physics in four
dimensions.  Unlike the determinant of the massless Schwinger model
[9], $\mbox{det}_{{Sch}}$ in (2.2) is not known explicitly.  Nevertheless
there are some important results.  One of these is an expression of
the paramagnetism of fermions in an external magnetic field as
summarized by the \lq\lq diamagnetic bound"
\begin{equation}
\mbox{det}_{{Sch}} \leq 1,
\end{equation}
for $m^2 \geq 0 [2, 10, 11]$.  This bound also expresses the positivity
of the one-loop effective action for Euclidean QED$_2$.  As a
consequence of (2.3) we were able to put an upper bound on the growth
of $\mbox{det}_{{ren}}$ in QED$_4$ for strong, unidirectional,
inhomogeneous magnetic fields.  Performing the dilation $A_\mu
\rightarrow \lambda A_\mu$ on the vector potential and letting
$\lambda$ become large we obtained

\begin{equation}
\ln \mbox{det}_{{ren}}\begin{array}{l}
\leq\\
{\lambda >> 1}
\end{array}
\displaystyle\frac{\lambda^2 e^2 V_{||} ||B ||^2}{24 \pi^2} \ln
\left(\displaystyle\frac{\lambda^2 e^2 || B||^2}{m^2}\right) +
0(\lambda^o),
\end{equation}
where $||B||^2 = \displaystyle\int d^2 r B^2 ({\bf r})$ [7].\\

There are other results for $\mbox{det}_{{Sch}}$ implicit in the
literature.  For example, if $A_\mu$ falls off sufficiently rapidly so
that $\displaystyle\int d^2r |A_\mu|^q < \infty$ for all $q \geq
\frac{1}{2}$, then one can relate $\mbox{det}_{{Sch}} (e)$ to the zeros
of the determinant, considered as a function of a complex coupling $e$
[12].  It is known [13] that for $m \neq 0$ these zeros occur in
quartets $e_n, -e_n, e^*_n$ and $-e^*_n$, and therefore
$\mbox{det}_{{Sch}}$ cannot vanish for real $e$ for these potentials
[14].\\

We will defer the discussion of the result of Haba [15] for
$\mbox{det}_{{ren}}$ until Sec.III.\\

In view of the direct connection between QED$_4$ and QED$_2$ in the
case of unidirectional magnetic fields it was thought worthwhile to
obtain more specific information on $\mbox{det}_{{Sch}}$ that would
enable one to make use of (2.1).  In a second paper [16] it was shown
that the exact calculation of $\mbox{det}_{{Sch}}$ reduces to a problem
in nonrelativistic, supersymmetric quantum mechanics.  That is to
say,
\begin{equation}
\ln \mbox{det}_{{Sch}} = \displaystyle\frac{e^2}{2\pi} \displaystyle\int
d^2 r \phi \partial^2 \phi + 2m^2 \displaystyle\int^e_0 d \lambda
Tr\{[(H_+ + m^2)^{-1} - (H_- + m^2)^{-1}]\phi\},
\end{equation}
where the supersymmetric operator pair $H_\pm = ({\bf p} - \lambda {\bf
A})^2 \mp \lambda B$ are obtained from the two-dimensional Pauli
Hamiltonian in (3.3) below.  The auxiliary potential, $\phi$, is
related to the vector potential by $A_\mu = \epsilon_{\mu \nu} \partial_\nu
\phi$ and to the magnetic field by $B = -\partial^2 \phi$ or

\begin{equation}
\phi ({\bf r}) = -\displaystyle\frac{1}{2\pi} \displaystyle\int d^2 r'
\ln |{\bf r} - {\bf r'}| B ({\bf r'}).
\end{equation}
The antisymmetric tensor $\epsilon_{\mu \nu}$ is normalized to
$\epsilon_{12} = 1$.  Again, the starting point for the derivation of
(2.5) was the proper time definition of $\mbox{det}_{{Sch}}$, which
respects gauge invariance and allows one to select the Lorentz gauge.
Our representation of $\mbox{det}_{{Sch}}$ makes a sharp separation
between the contribution from the massless Schwinger model, the first
term in (2.5), and its massive counterpart.  We have not integrated
the first term in (2.5) by parts as is usually done.  For nonzero flux
fields, $A_\mu$ falls off like $\frac{1}{r}$, and therefore an
integration by parts is invalid in this case.\\

Within the Lorentz gauge there is still the restricted gauge freedom
of $\phi \rightarrow \phi + c$, where $c$ is a constant.  Since
$\mbox{det}_{{Sch}}$ is gauge invariant, the term proportional to $c$ in
(2.5) must vanish, in which case

\begin{equation}
\displaystyle\frac{e^2 \Phi}{2\pi} = 2m^2 \displaystyle\int^e_0 d
\lambda Tr \left[(H_+ + m^2)^{-1} - (H_- + m^2)^{-1}\right],
\end{equation}
since $-\partial^2 \phi = B$ and $\Phi = \displaystyle\int d^2 rB$.
Differentiating (2.7) with respect to $e$ gives the index theorem on a
two-dimensional Euclidean manifold [17],\\

\begin{eqnarray}
\displaystyle\frac{e\Phi}{2\pi} &=& m^2 Tr\left[(H_+ + m^2)^{-1} - (H_- +
m^2)^{-1}\right]\nonumber\\
&=& n_+ - n_- + \displaystyle\frac{1}{\pi} \sum_l [\delta^l_+ (0) -
\delta ^l_- (0)],
\end{eqnarray}
where $n_\pm$ denote the number of positive and negative chirality
zero-energy bound states of $H_\pm$, and $\delta^l_\pm (0)$ are the
zero-energy phase shifts for scattering by the Hamiltonians $H_\pm$ in
a suitable angular momentum basis $l$.  Thus, the index theorem in
QED$_2$ follows from gauge invariance.\\

Now suppose (2.5) is written as

\begin{equation}
\ln \mbox{det}_{{Sch}} = \displaystyle\frac{e^2}{2\pi}\displaystyle\int
d^2 r \phi \partial^2 \phi + 2 \displaystyle\int^e_0 d \lambda Tr
\{[P_+^{(\lambda)} (0) - P_-^{(\lambda)} (0)]\phi\}
+\, {\mbox{nonzero modes,}}
\end{equation}
where $P_\pm^{(\lambda)} (0)$ are the zero-mode spectral measures
associated with $H_\pm$.  Noting that\\ $H_+ - H_- \sim B$ and assuming
$B$ sufficiently weak, we calculated the second term in (2.9) in first
Born approximation and showed that it cancelled the massless
Schwinger model term.  We conjectured that this was true in general,
and we will show that it is in Sec. IV.\\

Finally, we wish to retract a claim in [16].  By the Aharonov-Casher
theorem [18], $n_+(n_-)$ are given by
$\left[\displaystyle\frac{e|\Phi|}{2\pi}\right]$, depending on
whether $e \Phi > 0 (e \Phi < 0)$, where $[x]$ stands for the
nearest integer less than $x$ and [0] = 0.  Thus, if
$\displaystyle\frac{|e \Phi|}{2\pi} > 1$ there are zero-energy
bound states, and it was stated that when these are included in the
second term of (2.9) the logarithmic growth of $\phi$ and the slow,
algebraic fall off of the bound-state wave functions for large $r$
would cause $\mbox{det}_{{Sch}}$ to vanish.  This is false.  In fact, zero
modes are suppressed in $\mbox{det}_{{Sch}}$ as we will see in Sec.
IV.

\section{Choice of Fields}
\setcounter{equation}{0}

The fermion determinant is part of a functional integral whose measure,
$d \mu(A)$, is that of the gauge-fixed, free Maxwell field.  As
$d\mu(A)$ may be realized on $\cal{S}'$, the space of tempered distributions,
we seem to be stuck with rough potentials that are hard to analyze.  A
way out is to realize that there is a logarithmic ultraviolet
divergence in QED$_2$ due to the vacuum energy graph shown in Fig.1
that has to be regularized and subtracted out.  One way to regularize
[19] is to smooth $A_\mu$ by convoluting it with an ultraviolet cutoff
function $h_\Lambda \epsilon \cal{S}$, the functions of rapid decrease.
That is, let $A^\Lambda_\mu = A_\mu * h_\Lambda$ so that $A^\Lambda_\mu$ is
a polynomial bounded $C^\infty$ function, meaning that for each
$\alpha = (m,n)$ there is a $N(\alpha)$ and a $C(\alpha)$ with

\begin{equation}
\left| \displaystyle\frac{\partial^m}{\partial
x^m}\displaystyle\frac{\partial ^n}{\partial y^n} A^\Lambda_\mu (x,y)
\right| \leq C (1 + x^2 + y^2)^N.
\end{equation}
The regularized free photon propagator is
\begin{equation}
\displaystyle\int d \mu (A) A^\Lambda_\mu (x) A^\Lambda_\nu (y) =
D^\Lambda_{\mu \nu} (x -y),
\end{equation}
whose Fourier transform, $\hat{D}_{\mu \nu}^\Lambda (q)$, is proportional
to $|\hat{h}_\Lambda (q)|^2$, where $\hat{h}_\Lambda$ is the Fourier
transform of $h_\Lambda$.  One possibility is to choose $\hat{h}_\Lambda =
C^\infty_0$ with $\hat{h}_\Lambda (q) = 1$ for $q^2 \leq \Lambda^2$ and
$\hat{h}_\Lambda (q) = 0$ for $q^2 > 2 \Lambda^2$.  Note that the
measure $d \mu(A)$ is not regularized.  Thus, without loss of
generality we may assume that our potentials are smooth and polynomial
bounded.  Hereafter we will drop the superscript $\Lambda$ and denote
these potentials simply as $A_\mu$.\\

Now in order that the fermion determinant exist it seems necessary to
require the magnetic fields derived from the potentials to have finite
flux.  The reason for this restriction is connected with the
degeneracy of the ground state associated with the two-dimensional
Pauli Hamiltonian

\begin{equation}
H = ({\bf p} - {\bf A})^2 - \sigma^3 B \geq 0.
\end{equation}
Specifically, Avron and Seiler [20] have considered the class of
polynomial, infinite-flux magnetic fields

\begin{equation}
B({\bf r}) = \sum^N_{n=0} \lambda_n ({\bf r} - {\bf C}_n)^{2k_n},
\end{equation}
where $\{\lambda_n\}$ and $\{{\bf{C}}_n\}$ are arbitrary real numbers
and $\{k_n\}$ and N are nonnegative integers.  They have shown that
the ground state of H is infinitely degenerate and that the manifold
of zero-energy bound state wave functions is parametrized by a point
in ${\bf R}^{2(2k_{max} + 1)}$, irrespective of the translational
invariance of the magnetic field.  In the constant field case, $N =
0$, one has point spectrum, and the vector space is a plane whose
points specify the center of rotation of Landau orbits, corresponding to
the known degeneracy $|eB|L_xL_y /2\pi$, with $L_x, L_y \rightarrow
\infty$.  This degeneracy persists for all excited states in the
constant field case, and we suspect that the excited states for fields
with $N > 0$ are at least as degenerate as ${\bf R}^{2(2k_{max} + 1)}$.
Although we will show in Sec. IV that zero-modes are suppressed in our
representation (2.5) of $\mbox{det}_{{Sch}}$, it cannot make sense out of
the volume-like divergences associated with the degeneracy of the
excited modes.  Nor are we aware of any definition of determinant that
can.  Accordingly, we will confine our attention here to finite flux
fields.  In our view, a determinant of an infinite flux field should
be considered as the limit, if it exists, of a determinant calculated
from a field confined to local planar regions.\\

Restriction to finite flux is also consistent with the additional need
to regulate volume divergences before taking the thermodynamic limit
of the Euclidean Green's functions.  Such divergences appear in the
vacuum energy graphs, including the one in Fig.1, when the determinant
is integrated with respect to $d\mu(A)$.  A volume cutoff may be
introduced via the determinant, assumed previously calculated for a
generic class of fields, by replacing $B({\bf r})$ with $g({\bf r}) B
({\bf r})$, where $g$ is a suitable volume cutoff such as $g \epsilon
C^\infty_0$ or $g = \chi_\Lambda$, the characteristic function of a
bounded region $\Lambda \subset {\bf R}^2$.\\

At this point we mention the result of Haba [15], who studied
$\mbox{det}_{{ren}}$ in the field $F_{\mu \nu}(x) = G_{\mu \nu} +
F'_{\mu \nu} (x),$ where $G_{\mu\nu}$ is constant, and $F'_{\mu\nu}$
is sufficiently smooth and bounded.  It was concluded that with the
introduction of a space-time volume cutoff $\Lambda$ on the otherwise
infinite flux field $F_{\mu\nu}$ that

\begin{eqnarray*}
\ln \mbox{det}_{{ren}} &=& \displaystyle\frac{|\Lambda|}{48\pi^2} (G^2 -
\frac{3}{2} |G^*G|) \ln(1 + G^2)\\
&& -G^2 \displaystyle\int_\Lambda d^4x b (G, F'),
\end{eqnarray*}
where $b$ is bounded in $G$.  This saturates our upper bound (2.4) for
the special case $F'_{\mu\nu} = 0$ and $G_{12} = B$.  Haba then goes
on to use this result to obtain evidence suggesting the instability
(trivality) of QED$_4$.  In view of the potential importance of this
result and of our foregoing remarks concerning infinite flux fields,
it would be worthwhile to repeat the calculation with $F_{\mu\nu}$
confined to a finite region $\Lambda \subset {\bf R}^4$ from the
beginning and repeating Haba's estimates with $\Lambda$ held fixed.\\

As a consequence of the need to regularize and the finite flux
condition we may confine our attention to smooth, polynomial bounded
potentials that, in the Lorentz gauge, have the asymptotic form

\begin{equation}
A_\mu({\bf r}) = -\displaystyle\frac{\Phi}{2\pi}
\displaystyle\frac{\epsilon_{\mu \nu}x_\nu}{r^2} + 0(\frac{1}{r^2}),
\end{equation}
where $\Phi$ is the flux associated with $B$ and $r >> a$, the
range of $B$.  Our long-term goal is to study $\mbox{det}_{{Sch}}$ under
the scaling $B \rightarrow \lambda B, \lambda \rightarrow \infty$ as
well as the determinant's mass dependence.  This latter point, as we
now know, is especially relevant to determining $\ln \mbox{det}_{{ren}}$
in QED$_4$ for unidirectional, static, inhomogenous magnetic
fields.\\

We know of no previous calculation of a finite flux determinant
associated with massive fermions in two or more dimensions.  In Sec.
V
we will calculate such a determinant analytically for a magnetic field
confined to a thin cylindrical shell with radius {\it a}:\\

\begin{equation}
B(r) = \displaystyle\frac{\Phi}{2\pi} \displaystyle\frac{\delta (r -
a)}{a}.
\end{equation}
Although B is not derivable from a polynomial bounded potential it has
the virtue that the $\delta (r)/r$-type singularity of a
magnetic flux string is absent and that $\mbox{det}_{{Sch}}$ exists for
this field.  It is an instructive example, especially as we believe it
gives an insight to the matters raised above.

\section{Suppression Zero Modes in
$\mbox{\lowercase{d}\lowercase{e}\lowercase{t}}_{{S\lowercase{c}\lowercase{h}}}$}
\setcounter{equation}{0}

\subsection{Conventions}

Consider a Dirac fermion in a static, unidirectional magnetic field
directed along the $z$-axis.  Its Hamiltonian in the $xy$ plane is\\

\begin{equation}
H = \gamma^0 {\b \gamma} . ({\bf p} - e{\bf A}) + \gamma^0 m,
\end{equation}
with the $\gamma$ matrices\footnote{Where possible we adopt the notation and
conventions
 of
Jaroszewicz [21] whose analysis of the chiral anomaly associated with
the Hamiltonian (4.1) in a solenoidal magnetic field is relevant to
the work presented here.}$ \gamma^0 = \sigma^3, \gamma^k = -i
\sigma^k, k = 1,2.$  It has the structure
\begin{equation}
H = \left(
\begin{array}{cc}
m & L \\
L^{\dag} & -m
\end{array}
\right),
\end{equation}
where $L$ is a linear differential operator, and $L^{\dag}$ is the
Hermitian conjugate of $L$.  The positive- and negative-energy
eigenfunctions of $H$,

\begin{equation}
H \psi _{E, \lambda} ({\bf r}) = E \psi_{E,\lambda} ({\bf r}),
\end{equation}
are normalized to

\begin{equation}
\displaystyle\int d^2r \psi^{\dag}_{E, \lambda} ({\bf r})
\psi_{E',\lambda'} ({\bf r}) = \delta_{\lambda \lambda'} \delta (E -
E'),
\end{equation}
in the energy continuum, where $\lambda$ is a degeneracy parameter.
As a consequence of (4.2) the eigenvalues satisfy $E^2 \geq m^2$.
Since

\begin{equation}
H^2 = ({\bf p} - e{\bf A})^2 - eB \sigma^3 + m^2,
\end{equation}
we have

\begin{equation}
\left.
\begin{array}{c}
L  L^{\dag}\\
L^{\dag} L
\end{array}
\right\}
\qquad = ({\bf p} - e{\bf A})^2 \mp eB,
\end{equation}
so that the supersymmetric operator pair $H_\pm$ in Sec. III are given
by $H_+ = LL^{\dag}, H_- = L^{\dag} L$.  The eigenfunctions of $H_+$ and $H_-$
will be denoted by $u_{wl}$ and $v_{w,l},$ respectively,

\begin{eqnarray}
H_+ u_{w,l} ({\bf r}) &=& wu_{w,l} ({\bf r})\nonumber \\
H_- v_{w,l}(\bf r) &=& wv_{w,l} ({\bf r}),
\end{eqnarray}
where $l$ is a degeneracy paramater and $w = E^2 - m^2$.  Their normalization
in the
energy continuum, $w > 0$, is

\begin{equation}
\displaystyle\int d^2 r u^*_{w,l} ({\bf r}) u_{w',l'} ({\bf r}) =
\delta (w - w') \delta _{l l'}.
\end{equation}
Since the continuum extends down to $E = \pm m$ or $w = 0$ we expect
the states on the edge of the continuum to show up as unbounded
resonances [22].  When $|e\Phi|/2\pi > 1$, bound states will also
appear at the bottom of the continuum [18], and care must be taken to
include these in the completeness relations for $\psi_{E,\lambda}$ and
$u_{w,l,} \, v_{w,l}.$\\

\subsection{Suppression of Zero Modes}

Referring back to (2.5) consider

\begin{eqnarray}
\partial \ln \mbox{det}_{{Sch}}/\partial e &=& \displaystyle\frac{e}{\pi}
\displaystyle\int d^2 r \phi \partial^2 \phi\nonumber \\
&&+ 2m^2 \displaystyle\int d^2 r \phi ({\bf r}) \left\langle {\bf r}\left| (H_+
+ m^2)^{-1} - (H_- + m^2)^{-1}\right|{\bf r} \right\rangle.
\end{eqnarray}

Now consider the large mass limit of

\begin{eqnarray}
m^2 \left\langle {\bf r} \left| (H_+ + m^2)^{-1} - (H_- + m^2)^{-1} \right|{\bf
r}
\right\rangle  &=& m^2 \displaystyle\int^\infty_0dt e^{-tm^2} \left\langle {\bf
r} \left|
e^{-tH_+} - e^{-tH_-} \right|{\bf r} \right\rangle \nonumber \\
&=& m^2 \displaystyle\int^\infty_0 dt e^{- tm^2}
(\displaystyle\frac{e}{2\pi} B \left({\bf r}) + 0 (t) \right) \nonumber \\
&=& \displaystyle\frac{e}{2\pi} B ({\bf r}) + 0
\left(\displaystyle\frac{1}{m^2} \right),
\end{eqnarray}
where we used the heat kernel asymptotic expansion [16]

\begin{equation}
\left\langle {\bf r} \left| e^{-tH_\pm} \right|{\bf r} \right\rangle =
\displaystyle\frac{1}{4\pi t} \left[1 \pm teB ({\bf r}) + 0 (t^2) \right].
\end{equation}
Combining (4.7) and (4.10) we get

\begin{eqnarray}
\lefteqn{ \lim_{m^2 \rightarrow \infty} \left\langle {\bf r}\left|
\displaystyle\frac{m^2}{H_++
m^2} - \displaystyle\frac{m^2}{H_- + m^2}\right| {\bf r}
\right\rangle}\nonumber\\
&=& \lim_{m^2 \rightarrow \infty} \displaystyle\int^\infty_0 dw
\displaystyle\frac{m^2}{w + m^2} \sum_l (|u_{w,l} ({\bf r})|^2 -
|v_{w,l} ({\bf r})|^2)\nonumber\\
&=& \displaystyle\int^\infty_0 dw \sum_l (|u_{w,l} ({\bf r}) |^2 -
|v_{w,l} ({\bf r})|^2)\nonumber\\
&=& \displaystyle\int^\infty_0 \left\langle {\bf r}\left| d P_+ (w) -
d P_-(w) \right|{\bf r}
\right\rangle = \displaystyle\frac{e B({\bf r})}{2\pi},
\end{eqnarray}
where $P_+(w) - P_-(w)$ is the difference of spectral measures
associated with $H_\pm$ and where $H_\pm = \displaystyle\int^\infty_0 w d
P_\pm (w)$.\\

Equation (4.12) has the following physical interpretation.  The charge
density induced in the vacuum by the background magnetic field is

\begin{equation}
\langle j^0 ({\bf r)} \rangle = -\displaystyle\frac{e}{2}
\displaystyle\int^\infty_m dE \sum_\lambda (|\psi_{E,\lambda} ({\bf
r})|^2 - |\psi _{-E,\lambda} ({\bf r})|^2),
\end{equation}
which is determined by the spectral asymmetry density

\begin{eqnarray}
\eta(E,{\bf r}) &=& \displaystyle\frac{i}{2\pi} {\mbox{disc}}_E tr
\,\langle {\bf r}| (E - H)^{-1} - (E + H)^{-1} |{\bf r} \rangle\nonumber
\\
&=& \sum_\lambda (|\psi_{E,\lambda} ({\bf r})|^2 - |\psi_{-E,\lambda}
({\bf r})|^2),
\end{eqnarray}
where disc$_E h(E) \equiv h (E + i\epsilon) - h (E - i\epsilon)$.  Due
to the structure of H in (4.2), $\eta$ may also be expressed in terms
of the eigenfunctions of $H_\pm$ in (4.7) [21,23] as

\begin{equation}
\eta(E,{\bf r}) = 2m \sum_l (|u_{E^2 - m^2, l} ({\bf r})|^2 - |v_{E^2
- m^2,l} ({\bf r})|^2),
\end{equation}
so that

\begin{equation}
\langle j^0({\bf r}) \rangle = -\displaystyle\frac{e}{2}
\displaystyle\int^\infty_0 dw \displaystyle\frac{m}{\sqrt{w + m^2}}
\sum_l (|u_{w,l} ({\bf r})|^2 - |v_{w,l} ({\bf r})|^2).
\end{equation}
Comparing (4.16) with (4.12) we see that the limit in (4.12) is
equivalent to

\begin{equation}
\lim_{m \rightarrow \infty} \langle j^0 ({\bf r}) \rangle =
-\displaystyle\frac{e^2B({\bf r})}{4\pi}.
\end{equation}
The natural length scale here is the range of the magnetic field, and
so (4.17) states that the vacuum change density induced by a magnetic
field whose range is large compared to the fermion's Compton wave
length is $-e^2B({\bf r})/4\pi$, in agreement with a remark by
Jaroszewicz [21].\\

Now consider (4.9) again in the form

\begin{eqnarray}
\partial \ln \mbox{det}_{{Sch}}/\partial e &=& \displaystyle\frac{e}{\pi}
\displaystyle\int d^2r \phi \partial^2 \phi \nonumber \\
&&+ 2\displaystyle\int d^2 r \phi ({\bf r}) \displaystyle\int^\infty_0
dw \displaystyle\frac{m^2}{m^2 + w} \sum_l (|u_{w,l} ({\bf r}) |^2 -
|v_{w,l} ({\bf r}) |^2) \nonumber \\
&=& \displaystyle\frac{e}{\pi} \displaystyle\int d^2r \phi \partial^2
\phi\nonumber \\
&&+ 2\displaystyle\int d^2r\phi ({\bf r}) \displaystyle\int^\infty_0
(1 - \displaystyle\frac{w}{w + m^2}) \langle {\bf r}|dP_+(w) - dP_-
(w)|{\bf r}\rangle.
\end{eqnarray}
{}From the last line of (4.12) we finally obtain

\begin{eqnarray}
\partial \ln \mbox{det}_{{Sch}}/\partial e &=& -\displaystyle\frac{e}{\pi}
\displaystyle\int d^2 r \phi B + 2 \displaystyle\int d^2 r \phi (eB
/2\pi) \nonumber\\
&& -2 \displaystyle\int d^2 r \phi \displaystyle\int^\infty_0
\displaystyle\frac{w}{w + m^2} \langle {\bf r} | dP_+(w) - d
P_-(w)|{\bf r} \rangle  \nonumber\\
&=& -2 Tr \left[ \phi \left( \displaystyle\frac{H_+}{H_+ + m^2} -
\displaystyle\frac{H_-}{H_- + m^2} \right) \right].
\end{eqnarray}
which shows explicitly that zero modes are suppressed in
$\mbox{det}_{{Sch}}$.  It also shows how substantially mass has
altered the fermionic determinant of QED$_2$: the contribution of the
massless Schwinger model to $\mbox{det}_{{Sch}}$ has been cancelled by
a contribution from the massive sector of QED$_2$.

\subsection{Gauge Invariance of (4.19)}

If we revert back to the definitions (2.5) or (4.9) of
$\mbox{det}_{{Sch}}$, then the
invariance of $\mbox{det}_{{Sch}}$ under the restricted gauge
transformation $\phi \rightarrow \phi + c$ results in the index
theorem (2.8).  Recall that the massless Schwinger model's
contribution to $\mbox{det}_{{Sch}}$ was critical to establishing
this result.  Now, in (4.19), we see that the massless Schwinger
model's contribution has been cancelled by a contribution from the
massive sector of $\mbox{det}_{{Sch}}$ that contains the zero modes
of $H_\pm$.  Invariance of (4.19) under $\phi \rightarrow \phi + c$ now
requires that

\begin{equation}
\displaystyle\int d^2 r \displaystyle\int^\infty_0 dw
\displaystyle\frac{w}{w + m^2} \sum_l (|u_{w,l} ({\bf r})|^2 -
|v_{w,l} ({\bf r})|^2) = 0.
\end{equation}
In the remainder of this section we will recall the known result that
the integrand in (4.20) can be expressed as the divergence of a
current [21,23] and then go on to shown explicitly that (4.20) is true
in the case of radial symmetry.\\

Differentiating (4.7) with respect to $w$, denoted here by an overdot,
one easily obtains [21]

\begin{equation}
|u_{w,l} ({\bf r})|^2 - |v_{w,l} ({\bf r})|^2 = {\bf \nabla} {\b {^ .}} {\bf
S}_{w,l} ({\bf r}), \quad {\mbox for}\quad w > 0,
\end{equation}
where

\begin{equation}
{\bf S}_{w,l} = -[u^*_{w,l} {\bf {\nabla}} \dot{u}_{w,l} - ({\bf
{\nabla}} u^*_{w,l}) \dot{u}_{w,l} - 2iu^*_{w,l} {\bf A} \dot{u}_{w,l} -
(u \rightarrow v)].
\end{equation}
To make further progress we will specialize to the case of radial
symmetry so that the degeneracy paramater $l$ is identified with
angular momentum.  We would then like to interchange the space
integral with the sum over partial waves in (4.20) in order to convert
the space integral to an integral over a circle at infinity.
Jaroszewicz has already discussed this interchange in another context
[21], and we repeat his reasoning here.  Since $H_+ - H_- \sim B$, and
$B$ has finite range, $a$, the difference between the wave functions
$u_{w,l}$ and $v_{w,l}$, for fixed $w$, decreases with increasing $l$
due to the rising centrifugal barrier that excludes them from the
region where $B({\bf r}) \neq 0$.  The energy required for the wave
function to penetrate the region $r < a$ is of the order of $w >
\frac{l^2}{a^2}$, suggesting that the partial wave sum in (4.20) is
effectively cut off at $|l|\sim \sqrt{w}a$.  Accepting this reasoning,
we get
\begin{equation}
Tr\left(\displaystyle\frac{H_+}{H_+ + m^2} -
\displaystyle\frac{H_-}{H_- + m^2}\right) = \displaystyle\int^\infty_0
dw \displaystyle\frac{w}{w + m^2} \sum_l \lim_{R \rightarrow \infty}
\displaystyle\int_{S^1_R} {\bf S}_{w,l} ({\bf R}) {\b {^ .}} d{\bf l},
\end{equation}
where $d{\bf l} = {\bf \hat{R}} R d \theta$ in the $r, \theta$ plane.
Since the potential $\bf A$ may be assumed to be a pure gauge field
at infinity tangential to $S^1_\infty$, the $\bf A$-dependent terms in
(4.22) may be dropped.  Indeed our potentials (3.5) approach vortex
fields that manifestly satisfy this assumption.\\
Using the asymptotic form of $u_{w,l}$,
\begin{equation}
u_{w,l} (r, \theta)
\begin{array}[t]{c}
{\sim}\\[-5mm]
{\small r \gg a}
\end{array}
2^{-\frac{1}{2}} \pi^{-1}
w^{-\frac{1}{4}} r^{-\frac{1}{2}} e^{-il\theta} \cos\left[\sqrt{w} \,r -
\displaystyle\frac{\pi l}{2} - \displaystyle\frac{\pi}{4} + \delta^u_l
(w)\right],
\end{equation}
and similarly for $v_{w,l}$, where $\delta^u_l (w)$ and $\delta^v_l(w)$
are the scattering phase shifts, one gets [21]

\begin{equation}
\lim_{R \rightarrow \infty} \displaystyle\int_{S^1_R} {\bf S}_{w,l}
({\bf R}) {\b {^.}} d{\bf l} = \displaystyle\frac{1}{\pi} \sum_l
\left[\dot{\delta}^u_l (w) - \dot{\delta}^v_l(w)\right],\quad \mbox{for}
\quad w >
0,
\end{equation}
where the overdot continues to denote differentiation with respect to
$w$.  The factor $e^{-il\theta}$ instead of $e^{il\theta}$ in (4.24)
is for later notational convenience.  Due to the supersymmetry of the
operator pair $H_\pm$ we have from (4.7),

\begin{equation}
L^{\dag} H_+ u_{w,l} ({\bf r}) = H_- L^{\dag} u_{w,l} ({\bf r}) =  w L^{\dag}
u_{w,l} ({\bf r}),
\end{equation}
which indicates that $L^{\dag} u_{w,l} \propto v_{w,l}$ and hence that

\begin{equation}
\delta^{u}_{l}(w) = \delta^v_{l-1}(w),\, \mbox {mod}\, \pi
\, \mbox {for}\, w > 0,
\end{equation}
in agreement with Jaroszewicz [21].  In deriving (4.27) we used

\begin{eqnarray}
L^{\dag} &=& e^{i\theta} (\displaystyle\frac{\partial}{\partial r} - ieA_r +
\displaystyle\frac{i}{r} \displaystyle\frac{\partial}{\partial \theta} + e
A_\theta)\nonumber \\
& \begin{array}[t]{c}
     {\sim}\\[-5mm]
     {\small r \gg a}
     \end{array} &
                    e^{i \theta} (\displaystyle\frac{\partial}{\partial r} +
\displaystyle\frac{i}{r} \displaystyle\frac{\partial}{\partial \theta} +
\displaystyle\frac{e \Phi}{2\pi r}),
\end{eqnarray}
where ${\bf A} = (A_r, A_\theta)$.  The condition for the invariance
of (4.19) under the restricted gauge transformation $\phi \rightarrow
\phi + c$ is now reduced to
\begin{eqnarray}
\sum^\infty_{l = -\infty} \lbrack \dot{\delta}^u_l (w) - \dot{\delta}^v_l (w)
\rbrack &\equiv& \lim_{L \rightarrow \infty} \sum^L_{l = -L}
\lbrack \dot{\delta}^u_l (w) - \dot{\delta}^v_l (w)\rbrack \nonumber \\
&=& \lim_{L \rightarrow \infty} \lbrack \dot{\delta}^u_{-L} (w) -
\dot{\delta}^u_{L + 1} (w)\rbrack = 0,\quad \mbox {for}\quad w > 0.
\end{eqnarray}
This is physically reasonable since the wave equations for $u_{w,l}$ and
$v_{w,l}$
become scale invariant outside the range of $B$, where the potentials
$V_\pm$ defined by

\begin{equation}
H_\pm = -\partial^2 + V_\pm,
\end{equation}
have a $\frac{1}{r^2}$ behaviour and to where $u_{w,l}$ and $v_{w,l}$ are
mainly confined due to the enormous centrifugal barrier building up as
$|l| \rightarrow \infty$.  Hence, we do indeed expect $\lim_{L
\rightarrow \infty} \dot{\delta}^{u,v}_{|L|} (w) = 0$.\\

\section{$\mbox{\lowercase{d}\lowercase{e}\lowercase{t}}_{{S\lowercase{c}\lowercase{h}}}$ in a Cylindrical Magnetic Field}
\setcounter{equation}{0}

\subsection{The Green's Functions}

In this section we will use the representation (4.9) of $\mbox{det}_{{Sch}}$,
as it is simpler than (2.5) and more instructive than (4.19).  For the
magnetic field in (3.6) we have for the auxiliary potential $\phi$
given by (2.6),
\begin{eqnarray}
\phi (r) &=& -\displaystyle\frac{1}{2\pi} \displaystyle\int d^2r \ln
|{\bf r} - {\bf r'}| B(r')\nonumber \\
&=& \left\{  \begin{array}{ll}
               -\displaystyle\frac{\Phi}{2\pi} \ln r, & r > a \\[4mm]
               -\displaystyle\frac{\Phi}{2\pi} \ln a, & r < a.
	     \end{array}
     \right.
\end{eqnarray}
Since $\mbox{det}_{{Sch}}$ is invariant under $\phi \rightarrow \phi + c$
we may let $c = \displaystyle\frac{\Phi}{2\pi} \ln a$ and use the
potential
\begin{equation}
\phi(r) = -\displaystyle\frac{\Phi}{2\pi} \ln
\left(\displaystyle\frac{r}{a}\right) \theta (r - a),
\end{equation}
in which case the contribution of the massless Schwinger model is
eliminated straightaway:
\begin{eqnarray}
\displaystyle\int d^2r \phi \partial^2 \phi &=& -\displaystyle\int d^2 r
\phi B\nonumber \\
&=& -\displaystyle\frac{\Phi^2}{2\pi} \phi (a)\nonumber \\
&=& 0.
\end{eqnarray}
The associated vector potential is
\begin{eqnarray}
{\b A} &=& \displaystyle\frac{\Phi}{2\pi r} \theta (r-a) \b {\hat{\theta}}
\nonumber
\\
&\equiv& \displaystyle\frac{\Phi (r)}{r} \b {\hat{\theta}}.
\end{eqnarray}
Referring to (4.9) we see that the Green's functions $G_{\pm, l}$ defined
by

\begin{eqnarray}
\left\langle r, \theta |(k^2 - H_\pm)^{-1} |r', \theta' \right\rangle &=&
\displaystyle\frac{1}{2\pi} \sum^\infty_{l = -\infty} \left\langle r|(k^2 -
H_{\pm,l})^{-1}|r'\right\rangle e^{-il(\theta - \theta')}\nonumber \\
&=& \displaystyle\frac{1}{2\pi} \sum^\infty_{l = -\infty} G_{\pm, l}
(k; r, r') e^{-il(\theta - \theta')}.
\end{eqnarray}
are central to the calculation of $\mbox{det}_{{Sch}}$.  The radial wave
functions of $H_{+,l}$ and $H_{-,l}$, denoted by $u_{k^2, l} (r)$ and $v_{k^2,
l} (r)$, respectively, satisfy
\begin{eqnarray}
\left( -\displaystyle\frac{d^2}{dr^2} - \displaystyle\frac{1}{r}
\displaystyle\frac{d}{dr} + \displaystyle\frac{(l + e\Phi (r))^2}{r^2}
- \displaystyle\frac{e \Phi'(r)}{r}\right) u_{k^2,l} (r) &=& k^2
u_{k^2,l}(r)\nonumber \\
\left(-\displaystyle\frac{d^2}{dr^2} - \displaystyle\frac{1}{r}
\displaystyle\frac{d}{dr} + \displaystyle\frac{(l + e\Phi (r))^2}{r^2}
+ \displaystyle\frac{e\Phi'(r)}{r} \right) v_{k^2,l}(r) &=& k^2 v_{k^2,l} (r).
\end{eqnarray}
These equations have linearly independent solutions  $H^\pm_l(kr)$ for
$r < a$ and $H^\pm_{l + \Phi} (kr)$ for $r > a$, where $H^+_\nu$
and $H^-_\nu$ denote the Hankel functions $H^{(1)}_\nu$ and
$H^{(2)}_\nu$, respectively.  The dimensionless constant $e
\Phi/2\pi$ has been denoted simply by $\Phi$.

The calculation is simplified by introducing the Green's functions
\begin{equation}
{\cal G}_{\pm,l} (k;r,r') = \sqrt{rr'} G_{\pm,l} (k;r,r'),
\end{equation}
where
\begin{equation}
{\cal G}_{\pm,l} (k;r,r') = \langle r|(k^2 - {\cal H}_{\pm,l})^{-1}|r' \rangle,
\end{equation}
and
\begin{equation}
{\cal H}_{\pm,l} = -\displaystyle\frac{d^2}{dr^2} + \displaystyle\frac{(l +
\Phi(r))^2 - \frac{1}{4}}{r^2} \mp \displaystyle\frac{\Phi'(r)}{r},
\end{equation}
The outgoing-wave Green's functions ${\cal G}^+_{\xi,l} (\xi = \pm \equiv \pm
1)$ are constructed from [24]
\begin{equation}
{\cal G}^+_{\xi,l} (k;r,r') = -\displaystyle\frac{\phi (k,r_<)f_+
(k,r_>)}{{\cal J}_+ (k)},
\end{equation}
where $\phi$ and $f_+$ are regular and irregular solutions,
respectively, of
\begin{equation}
{\cal H}_{\xi,l} \psi_{\xi} = k^2 \psi_{\xi};
\end{equation}
${\cal J}_+$ for $\xi = \pm$ are the associated Jost functions, and $r_<,
r_>$ denote the lesser and larger values of $r, r'$, respectively.
The indices $l$ and $\xi$ have been omitted on the right-hand side of
(5.10) to reduce notational clutter.

Regular solutions of (5.11) are
\begin{eqnarray}
\phi(k,r) =
\left\{
\begin{array}{ll}
k^{-|l|} \sqrt{r} J_l (kr), & r < a,\\
k^{-|l|} \sqrt{r} \left[ \alpha H^+_{l + \Phi}  (kr) + \beta H^- _{l + \Phi}
(kr)\right], & r>a,
\end{array}
\right.
\end{eqnarray}
and the irregular, outgoing wave solutions are
\begin{eqnarray}
f_+ (k,r) &=& \left\{
\begin{array}{ll}
\sqrt{kr} \quad \left[ A H^+_l (kr) + BH^- _l (kr)
\right], & r <
a \\
\sqrt{kr} \quad H^+_{l + \Phi} (kr), & r > a.
\end{array}
\right.
\end{eqnarray}
The constants $\alpha, \beta, A, B$ are determined by the joining
conditions at $r = a$ obtained from (5.11) and (5.9):

\begin{eqnarray}
\phi (k, a-) &=& \phi (k, a+)\nonumber \\
\phi' (k,a+) - \phi' (k, a-) &=& -\xi \Phi \phi (k, a+)/a,
\end{eqnarray}
and similarly for $f_+$.  These give

\begin{eqnarray}
\alpha &=& \displaystyle\frac{i}{4} \pi \xi ka \left[J_l (ka) H^-_{l + \Phi
- \xi} (ka) - J_{l-\xi} (ka) H^-_{l + \Phi} (ka)\right]\nonumber \\
\beta &=& \displaystyle\frac{i}{4}\pi \xi ka \left[J_{l - \xi} (ka)
H^+_{l + \Phi} (ka) - J_l (ka) H^+_{l + \Phi -\xi} (ka)\right] \nonumber \\
A &=& \displaystyle\frac{i}{4}\pi \xi ka\left[H^+_{l + \Phi} (ka) H^-_{l
- \xi} (ka) - H^-_l (ka) H^+_{l + \Phi - \xi} (ka)\right] \nonumber \\
B &=& \displaystyle\frac{i}{4}\pi \xi ka \left[H^+_l (ka) H^+_{l + \Phi
- \xi} (ka) - H^+_{l - \xi} (ka) H^+_{l + \Phi} (ka)\right].
\end{eqnarray}
Finally, the Jost functions are
\begin{eqnarray}
{\cal J}_+ &=& W(f_+, \phi)\nonumber \\
&=& k^{-\frac{1}{2} - |l|} \beta r W[H^+_{l + \Phi} (kr), H^-_{l +
\Phi} (kr)]\nonumber \\
&=& -\displaystyle\frac{4i}{\pi} \beta k^{\frac{1}{2} - |l|},
\end{eqnarray}
where $W$ is the Wronskian, and $\beta$ is given in (5.15).\\

At this point ${\cal G}^+_{\pm,l}$ is fully determined by (5.10), (5.12),
(5.13), (5.15) and (5.16).  We could now continue ${\cal G}^+_{\pm,l}$ into
the upper half of the $k$ plane by letting $k = me^{i\theta}, \, \theta
\rightarrow \frac{\pi}{2}$ with $m > 0$, thereby making contact with
the Green's functions $\langle {\bf r} | (H_\pm + m^2)^{-1}| {\bf r'}\rangle$
in (4.9).  We will do this later.  But first we want to demonstrate
that when the flux is sufficiently large for fixed $l, {\cal G}^+_{\pm, l}$
does indeed contain a zero-energy bound state.  This will be done by
deriving a completeness relation.

\subsection{Completeness}

The proof of the completeness of the bound and scattering wave
functions associated with ${\cal H}_{\pm, l}$ will follow the general
procedure outlined by Newton [24], taking into account that
${\cal H}_{\pm,l}$ contains a $\frac{1}{r^2}$-type potential.\\

Consider the integral

\begin{eqnarray}
I(r) &=& \displaystyle\int_C dkk \displaystyle\int^\infty_0 dr' h(r')
{\cal G}^+_{\xi,l} (k;r,r')\nonumber \\
&=& -\displaystyle\int_C dkk \displaystyle\int^r_0 dr' h(r')
\displaystyle\frac{\phi (k,r') f_+ (k,r)}{{\cal J}_+ (k)}\nonumber \\
&&-\displaystyle\int_C dkk \displaystyle\int^\infty_r dr'
h(r')\displaystyle\frac{\phi (k,r) f_+ (k,r')}{{\cal J}_+ (k)}\nonumber \\
&=& I_1 + I_2,
\end{eqnarray}
where $h(r)$ is square-integrable, and $C$ is the contour in Fig.2.
The contribution of $I_{1\Gamma}$ to $I_1$ from the large semicircle
$\Gamma$ is evaluated by using the asymptotic forms of $\phi, f_+$
and $\cal {J}_+$ in the upper $k$ plane.  From (5.15) and (5.16),

\begin{eqnarray}
{\cal J}_+ &=& \xi a k^{\frac{3}{2} - |l|} \left[ J_{l - \xi} (ka) H^+_{l
+ \Phi} (ka) - J_l (ka)
H^+_{l + \Phi - \xi} (ka) \right] \nonumber \\
 & \begin{array}[t]{c}
     {\sim}\\[-5mm]
     {|k| \rightarrow \infty}
     \end{array} &
 - \displaystyle\frac{2i}{\pi}
k^{\frac{1}{2} - |l|} e^{- i \pi \Phi/2}.
\end{eqnarray}
{}From (5.12), (5.13) and (5.15) one gets, in the upper $k$ plane,
\begin{eqnarray}
\phi (k,r)
& \begin{array}[t]{c}
     {\sim}\\[-5mm]
     {|k| \rightarrow \infty}
     \end{array} &
\sqrt{\frac{2}{\pi}} k^{-|l| -
\frac{1}{2}} \cos(kr -\textstyle\frac{1}{4}\pi l -
\textstyle\frac{1}{4} \pi) \nonumber \\
f_+ (k,r)
& \begin{array}[t]{c}
     {\sim}\\[-5mm]
     {|k| \rightarrow \infty}
     \end{array} &
 \sqrt{\frac{2}{\pi}} e^{i(kr -
\textstyle\frac{1}{2} \pi (l + \Phi) - \textstyle\frac{1}{4}\pi)},
\end{eqnarray}
and hence for $|R|\rightarrow \infty$,
\begin{eqnarray}
I_{I\Gamma} &\sim& -i \displaystyle\int_\Gamma dk \displaystyle\int^r_0
dr'h(r') e^{i(kr - \frac{1}{2} \pi l - \frac{1}{4} \pi)} \cos (kr' -
\textstyle\frac{1}{2} \pi l - \textstyle\frac{1}{4}\pi)\nonumber \\
&\sim& -\displaystyle\frac{i}{2} h(r) \displaystyle\int_\Gamma dk
\displaystyle\int^r_0 dr' e^{ik(r-r')}\nonumber \\
&\sim& \displaystyle\frac{1}{2} i \pi h(r).
\end{eqnarray}
The contribution of $I_{2\Gamma}$ to $I_2$ from the contour $\Gamma$
may be defined by replacing the upper limit of the $r'$ integration in
(5.17) by $r + \mu, \mu > 0$ and letting $\mu \rightarrow \infty$
later [24].  Then for $|R| \rightarrow \infty$,

\begin{eqnarray}
I_{2\Gamma} &\sim& -i \displaystyle\int_\Gamma dk \displaystyle\int^{r
+ \mu}_r dr' h(r') e^{i(kr' - \frac{1}{2}\pi l - \frac{1}{4} \pi)}
\cos (kr - \textstyle\frac{1}{2} \pi l - \textstyle\frac{1}{4} \pi)\nonumber \\
&\sim& -\displaystyle\frac{i}{2} h(r) \displaystyle\int_\Gamma dk
\displaystyle\int^{r + \mu}_r dr' e^{ik(r' - r)} \nonumber \\
&\sim& \displaystyle\frac{1}{2} i\pi h(r).
\end{eqnarray}
Due to the analytic properties of $H^+_\nu (ka)$ and $J_l(ka)$ in
the upper half of the $k$ plane, ${\cal J}_+ (k)$ has no zeros for $Imk >0$.
For example, on the positive imaginary $k$-axis we have from (5.18),
\begin{equation}
{\cal J}_+(e^{\frac{i\pi}{2}} ka) = \displaystyle\frac{2iak}{\pi} e^{-i\pi(|l|
+ \Phi)/2} \left[ I_{l - \xi} (ka) K_{l + \Phi} (ka) + I_l (ka) K_{l +
\Phi - \xi} (ka)\right],
\end{equation}
where $K_\nu$ and $I_l$ are modified Bessel functions.  Since $K_\nu(x),
I_l(x)$ are positive for $x > 0,{\cal J}_+ (e^{i\pi/2} ka)$ is
manifestly free of zeros for $k > 0$.  Therefore,

\begin{equation}
\displaystyle\int_C dkk \displaystyle\int^\infty_0 dr' h (r')
{\cal G}^+_{\xi,l} (k; r, r') = 0.
\end{equation}
Combining (5.17) with (5.20), (5.21) and (5.23) we get
\begin{eqnarray}
h(r) &=& -\displaystyle\frac{i}{\pi} \displaystyle\int^{r + \mu}_{0} dr'
h(r') \left( \displaystyle\int^{-\epsilon}_{-\infty} +
\displaystyle\int^\infty_\epsilon \right) dkk \, \displaystyle\frac{\phi
(k,r_<) f_+(k,r_>)}{{\cal J}_+(k)} \nonumber\\ &&+ \displaystyle\frac{i}{\pi}
\displaystyle\int^{r + \mu}_{0} dr' h(r') \displaystyle\int_\gamma dkk
{\cal G}^+_{\xi,l} (k;r,r').
\end{eqnarray}
Equation (5.24) may be simplified by the following relations, valid
for positive $k$,

\begin{eqnarray}
\phi(ke^{i\pi},r) &=& \phi(k,r)\nonumber \\
f_+(ke^{i\pi},r) &=& -e^{i\pi/2} e^{-i\pi(l + \Phi)} f^*_+
(k,r)\nonumber \\
{\cal J}_+(ke^{i\pi}) &=& e^{3i\frac{\pi}{2}} e^{-i\pi|l|} e^{-i\pi \Phi}
{\cal J}_-(k),
\end{eqnarray}
where
\begin{eqnarray}
{\cal J}_-(k) &=& W(f_-, \phi)\nonumber \\
              &=& \displaystyle\frac{4i\alpha}{\pi} k^{\frac{1}{2} - |l|}\\
f_-(k,r)      &=& \sqrt{kr} H^-_{l + \Phi} (kr),\quad\quad r>a,
\end{eqnarray}
and where $\alpha$ is given by (5.15).  For real $k, {\cal J}^*_+(k) =
{\cal J}_-(k)$.  Equations (5.25), (5.26) and the result, valid for real $k$,

\begin{equation}
Im\left[{\cal J}_- (k) f_+ (k,r)\right] = \displaystyle\frac{2k}{\pi}
\phi(k,r), \quad r
< a,
\end{equation}
reduce (5.24) to
\begin{equation}
h(r) = \displaystyle\frac{4}{\pi^2} \displaystyle\int^{r + \mu}_0 dr'
h(r') \displaystyle\int^\infty_\epsilon dk
\displaystyle\frac{k^2}{|{\cal J}_+(k)|^2} \phi(k,r) \phi (k,r')\nonumber \\ +
\displaystyle\frac{i}{\pi} \displaystyle\int^{r + \mu}_0 dr' h(r')
\displaystyle\int_\gamma dkk {\cal G}^+_{\xi,l} (k;r,r').
\end{equation}
The integral around the semicircle $\gamma$ will contribute only if
${\cal G}^+_{\xi,l}$ develops a second-order pole at $k = 0$.  A tedious
calculation confirms that this happens for $\xi = +, l \leq 0, l +
\Phi > 1$ and for $\xi = -1, l \geq 0, l + \Phi < -1$.  This result
is in accord with a remark by Jaroszewicz [21], who stated this would
happen for $\Phi \neq$ integer; our result holds for all admissible
values of $\Phi$.  The residue of the double pole is the zero-energy
bound state $\psi_l$ of ${\cal H}_{\pm,l}$:
\begin{equation}
\lim_{\epsilon\downarrow 0} \displaystyle\int_\gamma dk k
{\cal G}^+_{\xi,l} (k;r;r') = -i\pi \psi_l (r) \psi_l (r'),
\end{equation}
for $\xi(l + \Phi) > 1, \xi l \leq 0$ and where
\begin{eqnarray}
\psi_l(r) &=& \left[ \displaystyle\frac{2(1 + |l|)(|l + \Phi|
-1)}{a^2|\Phi|} \right]^{\frac{1}{2}} \sqrt{r} \quad \quad
 \left\{
\begin{array}{ll}
\left(\displaystyle\frac{r}{a}\right)^{|l|}, & r < a  \\
\left( \displaystyle\frac{a}{r} \right)^{|l + \Phi|}, & r > a.
\end{array}
\right.
\end{eqnarray}
The bound states are normalized:

\begin{equation}
\displaystyle\int^\infty_0 dr \psi ^2_l (r) = 1.
\end{equation}
Combining (5.30) with (5.29) and letting $\mu \rightarrow \infty$
gives the completeness relation

\begin{equation}
\displaystyle\frac{4}{\pi^2} \displaystyle\int^\infty_{0^+} dk
\displaystyle\frac{k^2}{|{\cal J}_+ (k)|^2} \phi (k,r) \phi (k,r') +
\psi_l(r) \psi_l(r') = \delta (r-r').
\end{equation}

We have gone through this calculation to show that ${\cal G}^+_{\pm,l}$ does
indeed contain a bound state for sufficiently large flux and how it
manifests itself.  Later, when we consider the small mass limit of
$\mbox{det}_{{Sch}}$, one should keep in mind that this is being
controlled by the bound states, since the $m \rightarrow 0$ limit is
approaching the second-order pole of ${\cal G}^+_{\pm,l}$ along the positive
imaginary $k$-axis.

\subsection{Calculation of $\mbox{det}_{{Sch}}$}

The outgoing wave Green's function ${\cal G}^+_{\pm,l} (k; r, r')$ in
(5.10) may be continued to $k = im, m > 0$.  According to the definition of
$\mbox{det}_{{Sch}}$ in (4.9) we will only need
the difference of ${\cal G}^+_{+,l}$ and ${\cal G}^+_{-,l}$, which simplifies
the
calculation.  Due to (5.2), we may confine our attention to $r = r' >
a$.  Then from (5.10), (5.12), (5.13), (5.16) and (5.26),

\begin{equation}
{\cal G}^+_{+,l} (k;r,r) - {\cal G}^+_{-l} (k;r,r)\nonumber \\
= \displaystyle\frac{i\pi r}{4} \left( \left. \displaystyle\frac{{\cal J}_-
(k)}{{\cal J}_+(k)} \right|_{\xi =+} - \left. \displaystyle\frac{{\cal
J}_-(k)}{{\cal
 J}_+ (k)}
\right|_{\xi =-} \right)\quad (H^+_{l + \Phi} (kr))^2,
\end{equation}
which is related to the $S$-matrix for the $l$th partial wave by

\begin{equation}
S_l(k) = -\displaystyle\frac{{\cal J}_- (k)}{{\cal J}_+ (k)} e^{-i \pi \Phi} =
e^{2i
\delta_l}.
\end{equation}
In (5.34) the Jost ratios can be rearranged to give

\begin{equation}
\displaystyle\frac{{\cal J}_-}{{\cal J}_+} = 2 \displaystyle\frac{J_{l - \xi}
(ka)
J_{l + \Phi} (ka) - J_l(ka) J_{l + \Phi - \xi} (ka)}{J_{l - \xi}
(ka) H^+_{l + \Phi} (ka) - J_l (ka) H^+_{l + \Phi - \xi} (ka)} -1.
\end{equation}
We now let $k \rightarrow me^{i\frac{\pi}{2}}, m > 0$ in (5.34), (5.36)
and (5.8), using [25]

\begin{eqnarray}
J_\nu (am e^{i \pi/2}) &=& e^{i \pi \nu/2} I_\nu (ma)\nonumber
\\
H^+_\nu (am e^{i \pi/2}) &=& \displaystyle\frac{2}{\pi i}
e^{-i \pi \nu/2} K_\nu (ma),
\end{eqnarray}
to obtain

\begin{eqnarray}
\lefteqn{\left\langle r\left|({\cal H}_{+,l} + m^2)^{-1}\right|r \right\rangle
- \left\langle r\left|({\cal H}_{-,l} +
m^2)^{-1} \right|r \right\rangle }\nonumber \\
&=& -r \left( \displaystyle\frac{I_{l-1} I_{l + \Phi} - I_l I_{l + \Phi
-1}}{I_{l -1} K_{l + \Phi} + I_l K_{l + \Phi -1}} -
\displaystyle\frac{I_{l + 1} I_{l + \Phi} - I_l I_{l + \Phi + 1}}{I_{l
+ 1} K_{l + \Phi} + I_l K_{l + \Phi + 1}} \right) K^2_{l + \Phi} (mr),
\end{eqnarray}
where all modified Bessel functions in the brackets have argument $ma$.
If the two terms in (5.38) are combined by a common denominator then
a remarkable simplification occurs to give

\begin{eqnarray}
\lefteqn{
\left\langle r| ({\cal H}_{+,l} + m^2)^{-1} |r \right\rangle - \left\langle
r|({\cal H}_{-,l} +
m^2)^{-1} |r \right\rangle} \nonumber \\
&=& 2\Phi r \left\{ \lbrack ma \displaystyle\frac{d}{dma} \ln \left(
\displaystyle\frac{I_l (ma)}{K_{l + \Phi} (ma)} \right) \rbrack^2 -
\Phi^2 \right\}^{-1}\nonumber \\
&&\qquad\times
\left\{
\begin{array}{cc}
\displaystyle\frac{K^2_{l + \Phi} (mr)}{K^2_{l + \Phi} (ma)},
& r > a  \\[4mm]
\displaystyle\frac{I^2_l(mr)}{I^2_l(ma)}, & r < a,
\end{array}
\right.
\end{eqnarray}
where we have put on record the result for $r < a$.  As a check on
our results we verifed that

\begin{equation}
m^2 \displaystyle\int^\infty_0 dr \sum^\infty_{l = -\infty} \left[
\langle r\left|({\cal H}_{+,l} + m^2)^{-1}\right|r \right\rangle - \left\langle
r\left| ({\cal H}_{-,l} +
m^2)^{-1}\right|r \right\rangle = \Phi,
\end{equation}
for $m^2 > 0$, in accordance with (2.8).  This was done by
interchanging the integral and sum, which is allowed since the series
in (5.39) is uniformly convergent for $r \geq 0$, and then defining
$\sum^\infty_{l = -\infty}$ as $\lim_{L \rightarrow \infty}
\sum^L_{-L}$.  The integrals of the modified Bessel functions are
known [26].  Again, the reader is reminded that $\Phi$ denotes
$e\Phi/2\pi$, where $\Phi$ is the flux, and that this symbol is active
from (5.6) onward.

Finally, combining (4.9), (5.2), (5.5), (5.7) and (5.39) gives

\begin{eqnarray}
\displaystyle\frac{\partial \ln \mbox{det}_{{Sch}}}{\partial \Phi} &=&
-4(ma)^2 \Phi \displaystyle\int^\infty_1 dr r \ln r \sum^\infty_{l =
-\infty} \displaystyle\frac{K^2_{l + \Phi} (mar)}{K^2_{l + \Phi}
(ma)} \nonumber \\
&& \qquad\times \left\{ \left[ ma \displaystyle\frac{d}{dma} \ln \left(
\displaystyle\frac{I_l (ma)}{K_{l + \Phi} (ma)} \right) \right]^2 -
\Phi^2 \right\}^{-1}.
\end{eqnarray}
The term in the curly brackets is positive for all $\Phi$.  Figure 3
displays the numerical calculation of the right-hand side of (5.41)
for the cases $ma = 1$ and $10^{-2}$ for $0 \leq \Phi \leq 999.5$.
The plots were generated for half-integral values of $\Phi$.  The data
in both cases are consistent with a logarithmic growth of $\partial
\ln \mbox{det}_{{Sch}}/\partial \Phi$ with $\Phi$ given by $-\ln \Phi
+$ const, where the constant is about 0.3 for $ma = 1$ and $-2.5$ for
$ma = 10^{-2}$, consistent with the \lq\lq diamagnetic bound" in Eq.
(2.3).\\

The integral in (5.41) can be calculated explicity for integer $\Phi$.
Assume this is the case, and let $\Phi = N$.  Since the sum in (5.41)
is uniformly convergent for $r \geq 1$, we may let $l \rightarrow l-N$
and interchange sum and integral.  Since $K_{-l} = K_l$ and $I_{-l} =
I_l$, we need only
consider $l \geq 0$.  Then from [27] one can derive the result, valid
for $ l = 0, 1,\ldots$

\begin{eqnarray}
a_l &\equiv& \displaystyle\int^\infty_1 drr \ln r K^2_l \,(mar)/K^2_l
(ma)\nonumber \\
&=& -\displaystyle\frac{1}{2} \displaystyle\frac{d}{dma} \left(
\displaystyle\frac{K_{l + 1}}{K_l}\right) - \displaystyle\frac{l}{2(ma)^2}
\lbrack (-1)^l \displaystyle\frac{K^2_0}{K^2_l} + 2 \sum^l_{n =
1}(-1)^{l-n} \displaystyle\frac{K^2_n}{K^2_l} \rbrack,
\end{eqnarray}
where the modified Bessel functions on the right-hand side have argument $ma$.
This gives

\begin{eqnarray}
\left.\displaystyle\frac{\partial \ln \mbox{det}_{{Sch}}}{\partial \Phi}
\right|_{\Phi = N} &=& -2 (ma)^2 a_0 \left\{ \left[ ma
\displaystyle\frac{d}{dma} \ln \left( \displaystyle\frac{I_N}{K_0}
\right) - N \right] ^{-1}\right. \nonumber \\
&&\left.  - \left[ ma \displaystyle\frac{d}{dma} \ln \left(
\displaystyle\frac{I_N}{K_0} \right) + N \right] ^{-1} \right\}  \nonumber\\
&& - 2(ma)^2 \sum^\infty_{l = 1} a_l \left\{ \left[ ma
\displaystyle\frac{d}{dma} \ln \left( \displaystyle\frac{I_{l
-N}}{K_l} \right) - N \right] ^{-1}  \right.  \nonumber \\
&& + \left[ ma \displaystyle\frac{d}{dma} \ln \left(
\displaystyle\frac{I_{l + N}}{K_l} \right) - N \right] ^{-1}  \nonumber\\
&& - \left[ ma \displaystyle\frac{d}{dma} \ln \left(
\displaystyle\frac{I_{l -N}}{K_l}\right) + N \right] ^{-1}  \nonumber\\
&&\left.  - \left[ ma \displaystyle\frac{d}{dma} \ln \left(
\displaystyle\frac{I_{l + N}}{K_l}\right) + N \right] ^{-1} \right\} .
\end{eqnarray}

The remainder of this section will be confined to
an analysis of the asymptotic behaviour of $\mbox{det}_{{Sch}}$ for large
flux.  The case of small fermion mass is also of interest as this
limit will be controlled by the zero-energy bound states as noted in
Sec. V.B.  In addition, we conjecture that the low mass end of the
integral in (2.1) will control the large flux growth of
$\ln \mbox{det}_{{ren}}$.  Hence we are led to consider the limit $N
>> 1 >>ma$ of $\mbox{det}_{{Sch}}$.

\subsection{$\mbox{det}_{{Sch}}$ for $N >> 1 >> ma$}

The calculation of the above limit requires the large $l$ behaviour of
$a_l$ in (5.42).  Letting $x = ma$ and using

\begin{equation}
K_{l + 1} = \displaystyle\frac{2l}{x} K_l + K_{l -1},
\end{equation}
we get

\begin{equation}
a_l = -\displaystyle\frac{1}{2} \displaystyle\frac{d}{dx} \left(
\displaystyle\frac{K_{l -1}}{K_l} \right) + \displaystyle\frac{l}{x^2}
\left[ \left( \displaystyle\frac{K_{l-1}}{K_l} \right)^2 - \left(
\displaystyle\frac{K_{l-2}}{K_l} \right)^2 + \ldots +
\displaystyle\frac{1}{2} (-1)^{l+1} \displaystyle\frac{K^2_0}{K^2_l}
\right].
\end{equation}
{}From [28] one finds for $l >> 1>>x$

\begin{equation}
K_l(x) = \left( \displaystyle\frac{\pi}{2l} \right)^{\frac{1}{2}}
\left( \displaystyle\frac{2l}{ex}\right)^l \left[ 1 + \left(
\displaystyle\frac{1}{12} - \displaystyle\frac{1}{4} x^2 \right)
\displaystyle\frac{1}{l} + \left( \displaystyle\frac{1}{288} -
\displaystyle\frac{13}{48} x^2 + \displaystyle\frac{1}{32} x^4 \right)
\displaystyle\frac{1}{l^2}
+ 0 \left( \displaystyle\frac{1}{l^3} \right) \right],
\end{equation}
and hence

\begin{equation}
\displaystyle\frac{K_{l -1} (x)}{K_l (x)} = \displaystyle\frac{x}{2l}
\left[ 1 + \displaystyle\frac{1}{l} + \left( 1 -
\displaystyle\frac{1}{4} x^2 \right) \displaystyle\frac{1}{l^2} + 0
\left( \displaystyle\frac{1}{l^3} \right) \right],
\end{equation}
so that

\begin{equation}
a_l = \displaystyle\frac{1}{4l^2} + 0 \left(
\displaystyle\frac{1}{l^3} \right).
\end{equation}
Combining (5.48) with

\begin{equation}
x \displaystyle\frac{d}{dx} \ln \left( \displaystyle\frac{I_{l \pm
N}}{K_l} \right) > N,
\end{equation}
valid for $l, x \geq 0$, we may conclude that

\begin{equation}
\lim_{N \rightarrow \infty} \sum^\infty_{l=1} a_l \left[ x
\displaystyle\frac{d}{dx} \ln \left( \displaystyle\frac{I_{l \pm
N}}{K_l} \right) + N \right] ^{-1} = 0,
\end{equation}
in which case the second, fifth and last terms in (5.43) vanish in the
limit $N \rightarrow \infty$.\\

Considering the first term in (5.43) one finds that for $x <<1$,
\begin{equation}
x \frac{d}{dx} \ln \left( \frac{I_N}{K_0} \right) -
N = - \left[ \ln \left( \frac{x}{2} \right) \right] ^{-1} + 0
\left[ \ln \left( \frac{x}{2} \right) \right] ^{-2},
\end{equation}
and

\begin{equation}
x^2 a_0 = -\displaystyle\frac{1}{2} \left[ \ln \left(
\displaystyle\frac{x}{2} \right) \right]^{-1} + 0 \left[ \ln \left(
\displaystyle\frac{x}{2} \right) \right]^{-2},
\end{equation}
so that the first term in (5.43) tends to $-1$ for $N>>
1 >> x$.\\

Now consider the fourth term in (5.43).  Combining [28]

\begin{equation}
I_l(x) = (2\pi l)^{-\frac{1}{2}} \left( \displaystyle\frac{xe}{2l}
\right)^l \left( 1 + 0 \left( \displaystyle\frac{1}{l} \right) \right),
\end{equation}
for $l >> 1 >> x$ with (5.46) one finds

\begin{equation}
x \displaystyle\frac{d}{dx} \ln \left( \displaystyle\frac{I_{l +
N}}{K_l} \right) = 2l + N + 0 \left( \displaystyle\frac{1}{l}
\right),
\end{equation}
which together with (5.48) gives

\begin{equation}
\sum^\infty_{l = N+1} a_l \left[ x \displaystyle\frac{d}{dx} \ln
\left( \displaystyle\frac{I_{l + N}}{K_l} \right) - N \right] ^{-1} \sim
\displaystyle\frac{1}{8} \sum^\infty_{l = N + 1} l^{-3}.
\end{equation}
For the range $1 \leq l \leq N$, use (5.53) and $K'_l/K_l < 0$ for $l,
x, \geq 0$ to conclude, for $N >> 1, l \geq 1,$

\begin{equation}
x \displaystyle\frac{d}{dx} \ln \left( \displaystyle\frac{I_{l +
N}}{K_l} \right) - N \sim l - x \displaystyle\frac{K'_l}{K_l} > l.
\end{equation}
Therefore, for $x << 1$

\begin{equation}
\lim_{N \rightarrow \infty} \sum^\infty_{l = 1} a_l \left[ x
\displaystyle\frac{d}{dx} \ln \left( \displaystyle\frac{I_{l +
N}}{K_l} \right) - N \right]^{-1} = \quad \mbox {finite}.
\end{equation}

Finally, consider the remaining third term in (5.43).  For the range
$l \geq N + 1$ the third term may be written

\begin{equation}
\sum_{N + 1} \equiv \sum^\infty_{l = 1} a_{l + N} \left[ x
\displaystyle\frac{d}{dx} \ln \left( \displaystyle\frac{I_l}{K_{l + N}}
\right) - N \right]^{-1}.
\end{equation}
{}From (5.46) for $N >> 1 >> x, l \geq 1$

\begin{equation}
x \displaystyle\frac{d}{dx} \ln \left( \displaystyle\frac{I_l}{K_{l +
N}} \right) = x \displaystyle\frac{I'_l}{I_l} + l + N + 0 \left(
\displaystyle\frac{1}{N + l} \right) > l + N + 0 \left(
\displaystyle\frac{1}{N + l} \right),
\end{equation}
since $I'_l/I_l > 0$ for $l, x \geq 0$.  Equation (5.59) together with
(5.48) ensures that $\lim_{N \rightarrow \infty} \sum_{N + 1} = \,
\mbox{finite}$.

For the range $1 \leq l \leq N$ let

\begin{equation}
b_l = x^2 a_l \left[ x \displaystyle\frac{d}{dx} \ln \left(
\displaystyle\frac{I_{N - l}}{K_l} \right) - N \right] ^{-1},
\end{equation}
where we used $I_{-l} = I_l$.  We find for $x << 1$

\begin{equation}
b_1 = -\displaystyle\frac{1}{2} \ln x + 0(1)
\end{equation}

\begin{equation}
b_2 = \displaystyle\frac{1}{2} \left( 1 - \displaystyle\frac{1}{N}
\right) + 0 ([ x \ln x ]^2)
\end{equation}
The terms in (5.60) for $ 3 \leq l \leq N$ and $ x << 1$
behave as

\begin{equation}
a_l = \displaystyle\frac{1}{4} (l - 2)^{-2} -
\displaystyle\frac{x^2}{8} \displaystyle\frac{2l - 3}{(l - 1)^3 (l -
2)^3} + 0 (x^4),
\end{equation}

\begin{eqnarray}
x \displaystyle\frac{d}{dx} \ln \left( \displaystyle\frac{I_{N
-l}}{K_l} \right) - N &=& \displaystyle\frac{Nx^2}{2(N - l + 1) (l -
1)}\nonumber \\
&& + \displaystyle\frac{x^4}{8} \displaystyle\frac{3N^2l-N^3 - 4N^2 -
3l^2 N + 8Nl - 5N}{(l - 1)^2 (N - l + 1)^2 (l -2)(N - l + 2)}\nonumber
\\
&& + 0(x^6),
\end{eqnarray}
so that

\begin{equation}
\sum^N_{l = 3} b_l = \displaystyle\frac{1}{2} \ln N + 0(1).
\end{equation}
Combining the results obtained above on the sums in (5.43) together
with (5.61) and (5.65) give

\begin{equation}
\left. \frac{\partial \ln \mbox{det}_{{Sch}}}{\partial \Phi}
\right|_{\Phi = N} = -\ln \left( \displaystyle\frac{N}{ma} \right) +
0(1).
\end{equation}
The $-\ln N$ growth is consistent with the data in Fig.3.
Recall that $\Phi$ denotes $\displaystyle\frac{e \Phi}{2 \pi}$, the
two-dimensional chiral anomaly, and that $\left[\displaystyle\frac{e
\Phi}{2\pi}\right]$ is the number of zero-energy bound states of the
two-dimensional Pauli Hamiltonian (3.3).  Assuming a smooth variation
of $\mbox{det}_{{Sch}}$ with $\Phi$, we get, for $ma << 1
<< \displaystyle\frac{e \Phi}{2 \pi}$

\begin{equation}
\ln \mbox{det}_{{Sch}} = \quad - (\mbox {no. bound states}) X \quad \left[
\ln \left( \displaystyle\frac{\mbox {no. bound states}}{ma} \right) + 0
(1) \right].
\end{equation}
We suspect that (5.67) holds more generally than for the finite flux
magnetic field in (3.6).

\section{Zero Mass Limit of
$\mbox{\lowercase{d}\lowercase{e}\lowercase{t}}_{{S\lowercase{c}\lowercase{h}}}$}
\setcounter{equation}{0}

Consider either of the representations (2.5) or (4.19) for
$\mbox{det}_{{Sch}}$.  Does the $m = 0$ limit of $\mbox{det}_{{Sch}}$ exist?
If it does, is it continuous in the sense that the massless Schwinger
model's determinant is regained?  We do not have any definite answers
to these questions, but we suspect that they probably depend on
whether the magnetic fields have zero or nonzero flux as we will now
explain.\\

Already at the level of second-order perturbation theory one runs into
trouble for fields with $\Phi \neq 0$.  The determinant to order $e^2$
is

\begin{equation}
\ln \mbox{det}_{{Sch}} = -\displaystyle\frac{e^2}{2\pi}
\displaystyle\int \displaystyle\frac{d^2 q}{(2 \pi)^2} |\hat{B} (q)
|^2 \displaystyle\int^1_0 dz \displaystyle\frac{z(1 - z)}{z(1-z)q^2 +
m^2}.
\end{equation}
The finite flux field (3.6) gives

\begin{eqnarray}
\ln \mbox{det}_{{Sch}} &=& - \left( \displaystyle\frac{e \Phi}{2\pi}
\right)^2 \displaystyle\int^1_0 dz I_0 \left(
\displaystyle\frac{a|m|}{\sqrt{z(1-z)}} \right) K_0 \left(
\displaystyle\frac{a|m|}{\sqrt{z(1-z)}} \right)\nonumber \\
& \begin{array}[t]{c}
     {=}\\[-5mm]
     {m \rightarrow 0}
     \end{array} &
 - \left( \displaystyle\frac{e \Phi}{2\pi}
\right)^2 \ln (a|m|) + \mbox {finite},
\end{eqnarray}
while the zero flux field

\begin{equation}
B(r) = \displaystyle\frac{B}{r} \left[ \delta (r - a) - \delta (r -
b) \right],
\end{equation}
gives

\begin{equation}
\ln \mbox{det}_{{Sch}} = -(eB)^2 \displaystyle\int^1_0 d zz (1-z)
\displaystyle\int^\infty_0 d qq \displaystyle\frac{[J_0 (qa) -
J_0 (qb)]^2}{z(1-z)q^2 + m^2},
\end{equation}
which converges to the massless Schwinger model's determinant in the
limit $m = 0$:
\begin{eqnarray}
\ln \mbox{det}_{{Sch}} &=& -\displaystyle\frac{e^2}{2\pi}
\displaystyle\int \displaystyle\frac{d^2q}{(2\pi)^2}
\displaystyle\frac{|\hat{B} (q)|^2}{q^2}\nonumber \\
&=& -(eB)^2 \ln (a/b), \quad \mbox {for} \quad a > b > 0.
\end{eqnarray}

Going beyond perturbation theory we have from (5.67) for the finite flux
field (3.6)
\begin{equation}
\ln \mbox{det}_{{Sch}}
 \begin{array}[t]{c}
     {=}\\[-5mm]
     {m \rightarrow 0}\\
     {|e\Phi|>>|}
     \end{array}
\displaystyle\frac{|e\Phi|}{2\pi} \ln (|m|a) + \quad \mbox {finite at}\,
m = 0.
\end{equation}
This result agrees with intuition, that is

\begin{equation}
\ln \left[ \displaystyle\frac{\det (H + m^2)}{\det (p^2 + m^2)}
\right]^{\frac{1}{2}}
 \begin{array}[t]{c}
     {=}\\[-5mm]
     {m \rightarrow 0}
     \end{array}
\quad (\mbox{no. zero modes
of H}) \times \ln (|m| R) + \mbox {finite},
\end{equation}
where $H$ is the Pauli Hamiltonian (3.3), and $R$ is some natural
length scale, such as the range of $B$.  The massless Schwinger
model's determinant is not regained.\\

Finally, Seiler [12] defined the massless Schwinger model's
determinant, $\det (1 - e \frac{1}{\pslash} \Aslash)$, by a renormalized
determinant.  This was done by making a formal similarity
transformation and defining the determinant as $\mbox{det}_{{ren}} (1 -
eK)$ with
\begin{equation}
K(A) = \displaystyle\frac{\pslash}{|p|^{\frac{3}{2}}} \Aslash
\displaystyle\frac{1}{|p|^{\frac{1}{2}}},
\end{equation}
where $K$ is considered as an operator on two-component
square-integrable functions.  The determinant $\mbox{det}_{{ren}}$
excludes the linearly divergent graph $TrK$ but retains the graph
$TrK^2$, which is defined in some gauge invariant way.  Assuming that
the magnetic fields have zero flux and in particular that
$\displaystyle\int d^2r |A_\mu|^q < \infty$ for all $q \geq
\frac{1}{2}$, Seiler was able to show that $K \in {\cal C}_n$ (the space of
operators for which $Tr(K^{\dag} K)^{\frac{n}{2}} < \infty$) for all $n
> 2$ and that the spectrum of $K$ consists only of the origin.  This
latter result is an expression of the triviality of the massless
Schwinger model and implies that all single-loop \lq\lq photon-photon"
scattering graphs of order $e^4$ and higher vanish [29].  These results
allowed Seiler to obtain the well-known result for the massless
Schwinger model's determinant
\begin{equation}
\ln \mbox{det}_{{ren}} (1 - eK) = -\displaystyle\frac{e^2}{2\pi}
\displaystyle\int d^2 r A^2_\mu,
\end{equation}
where $A_\mu$ is in the Lorentz gauge.\\

In the nonzero flux case $A_\mu$ falls off as $\frac{1}{r}$, and
$K(A)$ ceases to be even a compact operator:  the eigenvalues of $K$
fill an open disc and $K \notin {\cal C}_n$ for any $n \geq 1$.  Consequently,
it is no longer possible to define the massless Schwinger model's
determinant in terms of a renormalized determinant whose zeros
reflect the eigenvalues of $K(A)$.\\

The lesson here is that the transition from a zero-flux magnetic field
to one with nonzero flux is not a smooth one and that the zero-mass
limit of $\mbox{det}_{{Sch}}$ will be flux sensitive.

\section{NET RESULT}
\setcounter{equation}{0}

Fermionic determinants are at the heart of fermionic field theories.
In the case of QED the determinant in 3 + 1 dimensions in a static,
unidirectional, finite (including zero) flux magnetic field can be
calculated from the determinant of the massive Euclidean Schwinger
model, $\mbox{det}_{{Sch}}$, in the same magnetic field by integrating over the
fermion's mass.  Therefore, the massive Schwinger model is physically
relevant.  The calculation of $\mbox{det}_{{Sch}}$ reduces to a problem in
nonrelativistic, supersymmetric quantum mechanics, and the gauge
invariance of $\mbox{det}_{{Sch}}$ is closely linked to the index theorem on a
two-dimensional Euclidean manifold.  The inclusion of mass
qualitatively changes the determinant in 1 + 1 dimensions to the
extent that the massless Schwinger model's contribution to $\mbox{det}_{{Sch}}$
is cancelled by a contribution from the massive sector.  Evidence was
given that the zero-mass limit of $\mbox{det}_{{Sch}}$ is not continuous in the
sense that the massless Schwinger model's determinant is not regained
for nonzero flux magnetic fields.\\

It is believed that the first calculation of $\mbox{det}_{{Sch}}$ for a finite
flux magnetic field is given in Sec.V.  The behaviour of the
determinant for large flux and small mass suggests that the
zero-energy bound states of the two-dimensional Pauli Hamiltonian are
the controlling factor in the growth of $\ln \mbox{det}_{{Sch}}$.  If this is
the case then the implication of this fact on the still unknown growth
of the renormalized determinant of QED$_4$ in the same magnetic field,
which is determined by (2.1), remains to be seen.\\

\vspace*{4mm}

\begin{center}
\bf Acknowledgements
\end{center}

\vspace*{4mm}

The author wishes to thank his colleagues for useful discussions,
especially D. J. O'Connor.  He also wishes to thank R. Russell for
performing the numerical calculations displayed in Fig.3.

\newpage

\begin{center}
FIGURE CAPTIONS
\end{center}

\vspace*{6mm}

\begin{tabular}{ll}
FIG. 1. &  The logarithmically divergent contribution to the vacuum
energy in QED$_2$.\\[6mm]
FIG. 2.  & The contour of integration, C, in the $k$-plane for the
integral in (5.17).\\[6mm]
FIG. 3.  & Numerical calculation of the right-hand side of Eq. (5.41)
for half-integral\\
& values of $\Phi$ with $ma = 1$ and $10^{-2}$.
\end{tabular}

\end{document}